\documentclass[aps,prl,reprint]{revtex4-1}

\usepackage[caption=false]{subfig}
\usepackage[colorlinks=true, citecolor=blue, linkcolor=blue, urlcolor=blue]{hyperref}
\usepackage{graphicx}
\usepackage{amsmath}
\usepackage{makecell}
\usepackage{soul}
\usepackage{soul,color,xcolor}
\usepackage{framed}
\usepackage{bm}
\usepackage[listings,breakable,xparse]{tcolorbox}
\newcommand{\figsize}{0.48}
\definecolor{greenshade}{rgb}{0.90,0.99,0.91}
\sethlcolor{greenshade}
\newcommand{\prlhead}[1]{\textit{#1}\kern0pt---\kern0pt\ignorespaces}

\newcommand{\prlshead}[1]{#1:\kern1em\ignorespaces}

\newcommand{\prlsshead}[1]{\textit{#1.}\kern1em\ignorespaces}

\newcommand{\prlstep}[1]{#1:\kern1em\ignorespaces}

\newcommand{\prltheorem}[1]{\textit{#1}\kern0pt---\kern0pt\ignorespaces}

\newcommand{\prllemma}[1]{\textit{#1}\kern0pt---\kern0pt\ignorespaces}

\begin{document}


\title{ Square-Root  Higher-Order Exceptional Points with Symmetry-Induced Multiple Spectral Responses}


\author{Haoyang Zhang$^{1,5}$}
\author{Yadi Niu$^{1}$}
\author{Nuo Wang$^{1}$}
\author{Zihan Mo$^{1}$}
\author{Ying Gu$^{1,2,3,4,5,}$}
\email[]{ygu@pku.edu.cn}
\affiliation{$^1$State Key Laboratory of Artificial Microstructure and Mesoscopic Physics $\&$ Department of Physics, Peking University, Beijing 100871, China\\ $^2$Frontiers Science Center for Nano-optoelectronics $\&$  Collaborative Innovation Center of Quantum Matter, Peking University, Beijing 100871, China\\$^3$Collaborative Innovation Center of Extreme Optics, Shanxi University, Taiyuan, Shanxi 030006, China\\$^4$Peking University Yangtze Delta Institute of Optoelectronics, Nantong 226010, China\\$^5$Hefei National Laboratory, Hefei 230088, China}


\date{\today}

\begin{abstract}
We generalize square-root procedure to non-Hermitian systems with finite lattices, providing a spectral operation scheme applicable to arbitrary tight-binding models.
Via this generalized square-root approach, we construct novel      
chiral-symmetric higher-order exceptional points (EPs) with multiple spectral responses.
By taking square-root of a parent Hamiltonian hosting an $n$th-order EP (EP$_n$), an EP$_{2n+1}$ chiral-symmetric square-root system is obtained, whose lattice sites are inherited from both the parent system and an auxiliary residual system.
The chiral-symmetry-induced structure of the generalized eigenspace enables onsite and coupling perturbations to selectively generate spectral responses of different orders.
The proposed scheme is universal, applicable to any existing tight-binding EP system and iterable to generate EPs of arbitrarily high order.
With enriched ultrasensitive spectral responses, the chiral-symmetric higher-order EP system provides a promising platform for signal amplification, detection and non-Hermitian control.
\end{abstract}


\maketitle

\prlhead{Introduction.}The complex energy spectrum in non-Hermitian systems exhibits high sensitivity to small parameter perturbations.
This sensitivity enables ultrasensitive measurements beyond the Hermitian linear-response limit and provides new physical principles for enhancing weak signals and light–matter interactions   \cite{2014prlEnhancing,2016praSensors,2020prReview}.
At its core, such sensitivity originates from a distinctive type of degeneracy known as exceptional points (EPs).
At an EP of order $n$ (EP$_n$), $n$ eigenvectors coalesce into a single state, and this degeneracy is readily lifted by an infinitesimal perturbation.
Specifically, near an EP$_n$, an observable (e.g., frequency splitting) scales as the $n$-th root of the perturbation strength $\epsilon$, thereby yielding a nonlinear amplification factor of $\epsilon^{-1+1/n}\gg1$  \cite{1995springerPerturbation}.
Therefore, higher-order exceptional points (HOEPs) enable stronger sensitivity enhancement and weak-signal amplification  \cite{2012jpmSignatures,2022prrResponse}.
EP$_3$, at which three eigenvalues and their eigenstates coalesce, is experimentally demonstrated to yield stronger spectral sensitivity than EP$_2$ \cite{2017natureEnhanced}. EPs of even higher order amplify the frequency splitting for ultrahigh-precision sensing \cite{2019praHigh-order,2023prHigher-order}.
Beyond sensing, HOEPs also show potential in spontaneous emission enhancement \cite{2016prlEnhanced}, cooling in optomechanical cavities \cite{2017srHigh-order}, flat-band filters \cite{2020prafilters}, and fast entanglement generation \cite{2023prlSpeeding}.

The universal and flexible realization of HOEPs remains a key challenge in non-Hermitian research.
HOEPs up to order 3,4, and 6 have been realized through carefully designed coupling structures \cite{2016prxEmergence,2020prawaveguide-induced,2025srexceptional}. However, these designs rely on a limited set of known coupling configurations, thus cannot be straightforwardly extended to arbitrary high order.
Alternatively, via unidirectional coupling, HOEPs can be obtained by hierarchically stacking two EP systems \cite{2020prlhierarchical,2022prarevisiting,2025prrcomposite}, causing their two sets of eigenenergy spectra to coalesce in a nontrivial manner.
But, perfect unidirectional coupling poses difficulties for optical implementation.
Recently, as a general framework, symmetry is introduced in periodic non-Hermitian systems to relax the  requirements of HOEPs, where different symmetry classes yield qualitatively distinct types of EPs \cite{2021prlSymmetry,2022nrpNon-hermitian}.
Nevertheless, in these periodic systems, the order of EPs is not very high, and also their band symmetries cannot be directly applied to finite lattices.
Therefore, symmetry-enabled exceptional points of arbitrarily high order have yet to be achieved in finite-lattice systems.

\begin{figure}[t]
	\includegraphics[width=\figsize\textwidth]{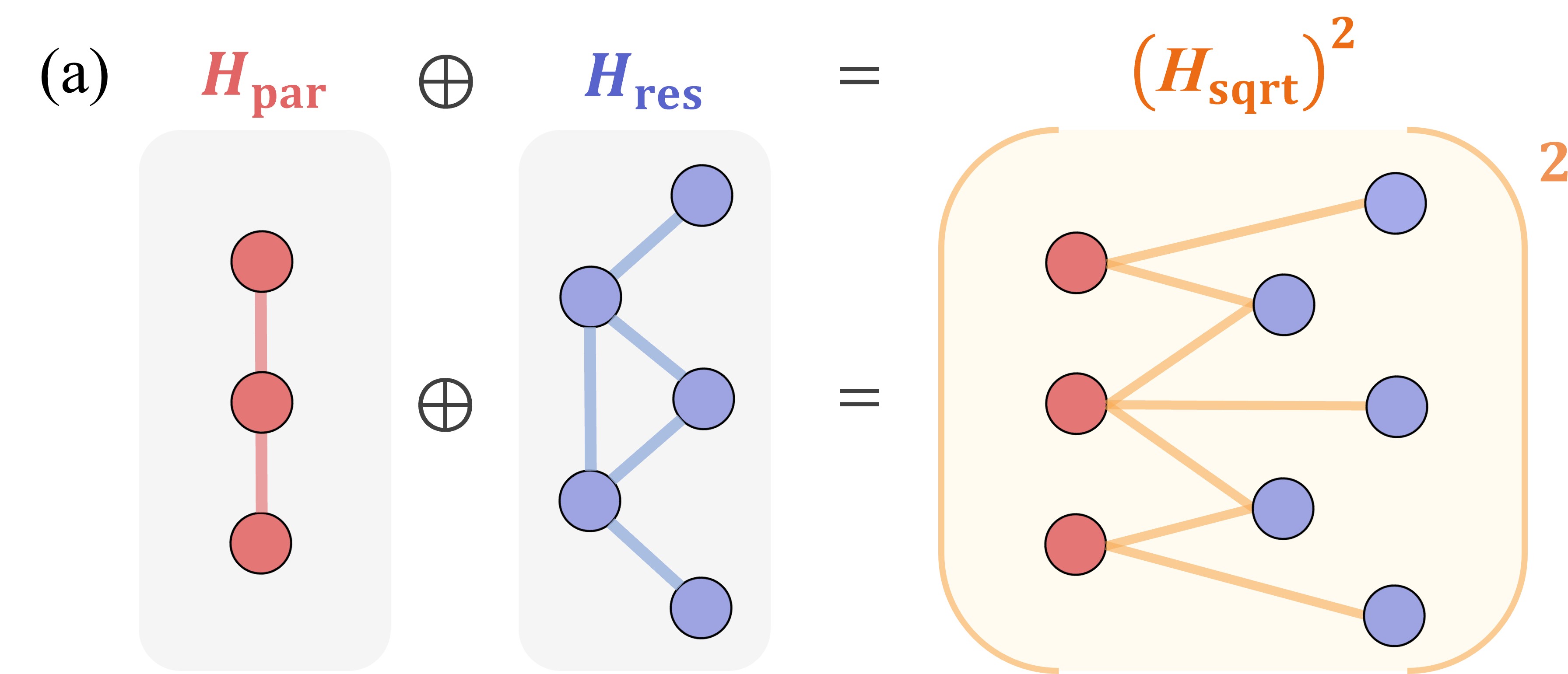}
	\includegraphics[width=\figsize\textwidth]{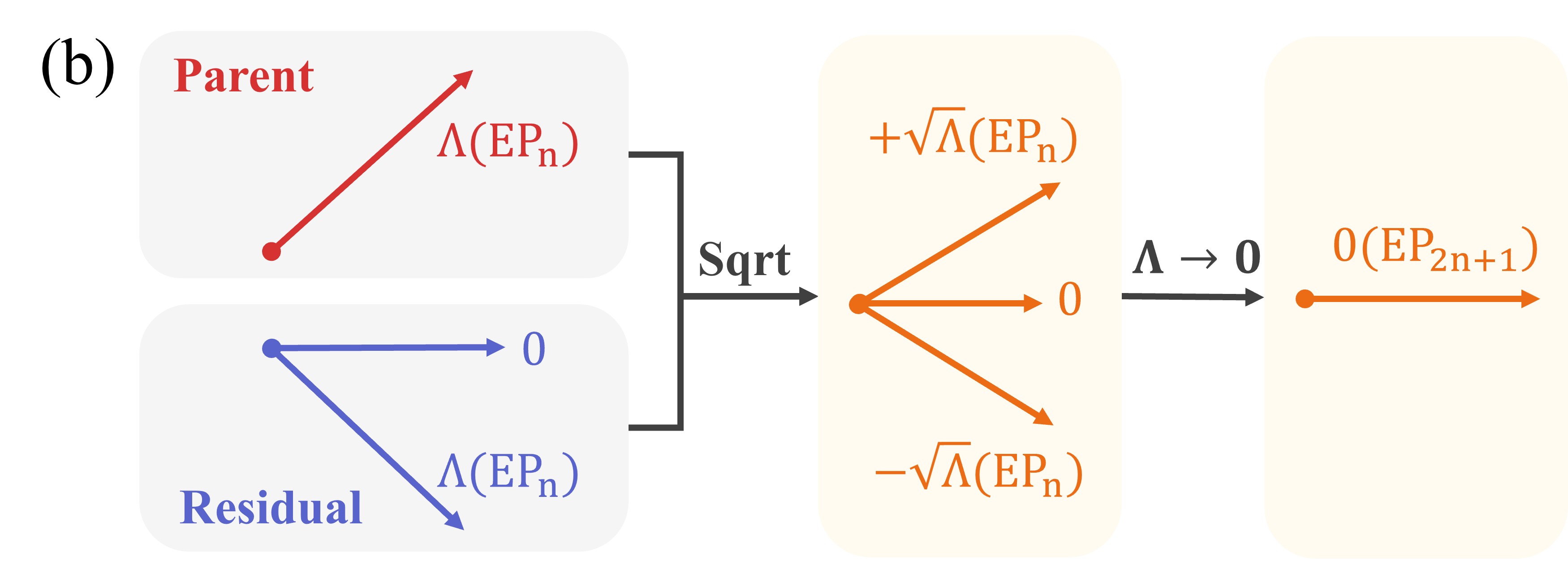}
	\caption{\label{fig:1} Schematic diagram of generating HOEPs via a square-root process.
	 (a) Construction of square-root Hamiltonian from parent and residual systems, and (b) corresponding eigenvalues and eigenstates.} 
\end{figure}

Here, we generalize square-root procedure \cite{2017prbnontrivial,2020prrSystematic} to non-Hermitian systems with finite lattices, providing a spectral operation scheme applicable to arbitrary tight-binding models. Via this generalized method, chiral symmetry can be introduced into the non-Hermitian parent system, serves as the key ingredient for realizing the higher order of EPs.
Specifically, starting from a parent system hosting an EP$_n$, we construct its square-root Hamiltonian, whose lattice sites originate from both the parent and an auxiliary
residual system [Fig.~\ref{fig:1}(a)].
The square-root operation intrinsically introduces chiral symmetry, which splits the parent EP$_n$ into two branches.
Crucially, when parent EP$_n$ has zero eigenvalue, the eigenstates of two split EP$_n$'s also coalesce, yielding an EP$_{2n+1}$ square-root system [Fig.~\ref{fig:1}(b)].
This universal framework can be applied to any tight-binding system and iterated to generate arbitrarily high-order EP. 
Moreover, perturbations applied to on-site or coupling terms can selectively break chiral symmetry of this HOEP system, thereby generating different order spectral responses in a controlled manner.
With enriched ultrasensitive spectral responses, the chiral-symmetric HOEP system provides a promising platform for signal amplification, detection and non-Hermitian control.


\prlhead{Square-root construction of ${\rm HOEPs}$.}
The square-root procedure constructs the system whose eigenvalues are the square roots of those of the parent Hamiltonian, thereby doubling both the energy bands and band gaps \cite{2017prbnontrivial,2020prrSystematic}. 
When applying to topological periodic systems, 
chiral symmetry introduced by the square-root process gives rise to higher-order topological insulators \cite{2021prb2nsemimetals,2021prbsuperconductors,2021acsRealization} and novel topological phases \cite{2021prbSquare-root,2022ncweyl,2022prbcrystals}.
It has been experimentally realized in optical waveguides \cite{2021acsRealization,2019lprlattice,2023lprArrays}, acoustic systems \cite{2020prbAcoustic} and electronic circuits \cite{2020nlCircuits,2023aplCircuits}. 
Recently, generalized nth-root operations are applied to non-Hermitian topological systems \cite{2024Nanonroot,2026prbhorns}, yielding distinctive spectrum structures like exceptional horns.
Here, we generalize this procedure to finite-sized non-Hermitian tight-binding lattices, then the eigenvalue spectrum precisely splits into two square-root branches.

For a given parent system, the Hamiltonian of the square-root system is constructed by coupling the parent lattice sites to residual lattice sites [Fig.~\ref{fig:1}(a)] 
\begin{equation}
	{H}_{\rm sqrt}=\left[\begin{array}{cc}{\textbf{0}}& {H}^{\rm left}\\ {H}^{\rm right}& {\textbf{0}}\end{array}\right],
\end{equation}
where \(H^{\rm left}\) and \(H^{\rm right}\) describe the residual-to-parent and parent-to-residual couplings, respectively.
The residual sites serve two distinct functions in the construction.
One  reproduces the desired coupling  of the parent Hamiltonian, while the other is to engineer the effective on-site energies (The details are shown in Ref. \cite{SM}).
Accordingly,  the parent Hamiltonian can exactly recover through the coupling, $H_{\rm par} = H^{\rm left}H^{\rm right}$, i.e., 
\begin{equation}
	{H}_{\rm sqrt}^{2}
		=\left[\begin{array}{cc}{H}^{\rm left}{H}^{\rm right}& {\textbf{0}}\\ {\textbf{0}}& {H}^{\rm right}{H}^{\rm left}\end{array}\right]
		=\left[\begin{array}{cc}{H}_{\rm par}& {\textbf{0}}\\ {\textbf{0}}& {H}_{\rm res}\end{array}\right],
\label{eq:2}
\end{equation}
with the residual Hamiltonian $H_{\rm res} = H^{\rm right}H^{\rm left}$.
Please note that the above procedure can  serve as a spectral operation scheme applicable to arbitrary tight-binding models.
In the square-root process, by appropriately adjusting the effective on-site energies, the zero-energy reference points of parent system can be freely set. By satisfying Eq.~\eqref{eq:2},  the spectrum in zero-energy exactly follows the square-root relation, enabling the further coalescence of square-root eigenstates into higher order of EPs  [Fig.~\ref{fig:1}(b)].
While in periodic systems, the square-root constructions shift the parent spectrum away from zero \cite{2017prbnontrivial,2020prrSystematic}, preventing the generation of higher order of EPs.

\begin{figure*}
	\centering
	\subfloat{\includegraphics[width=\figsize\textwidth]{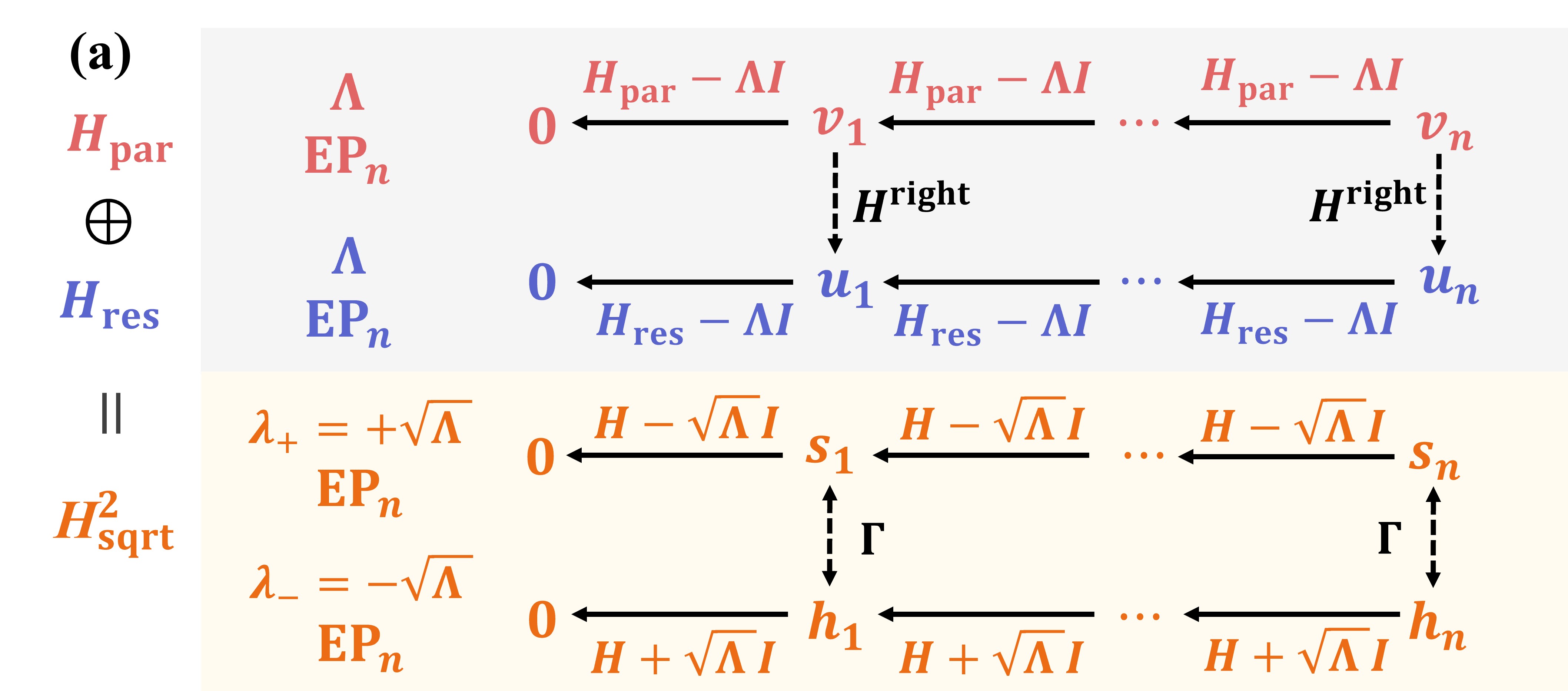}}\hfill
	\subfloat{\includegraphics[width=\figsize\textwidth]{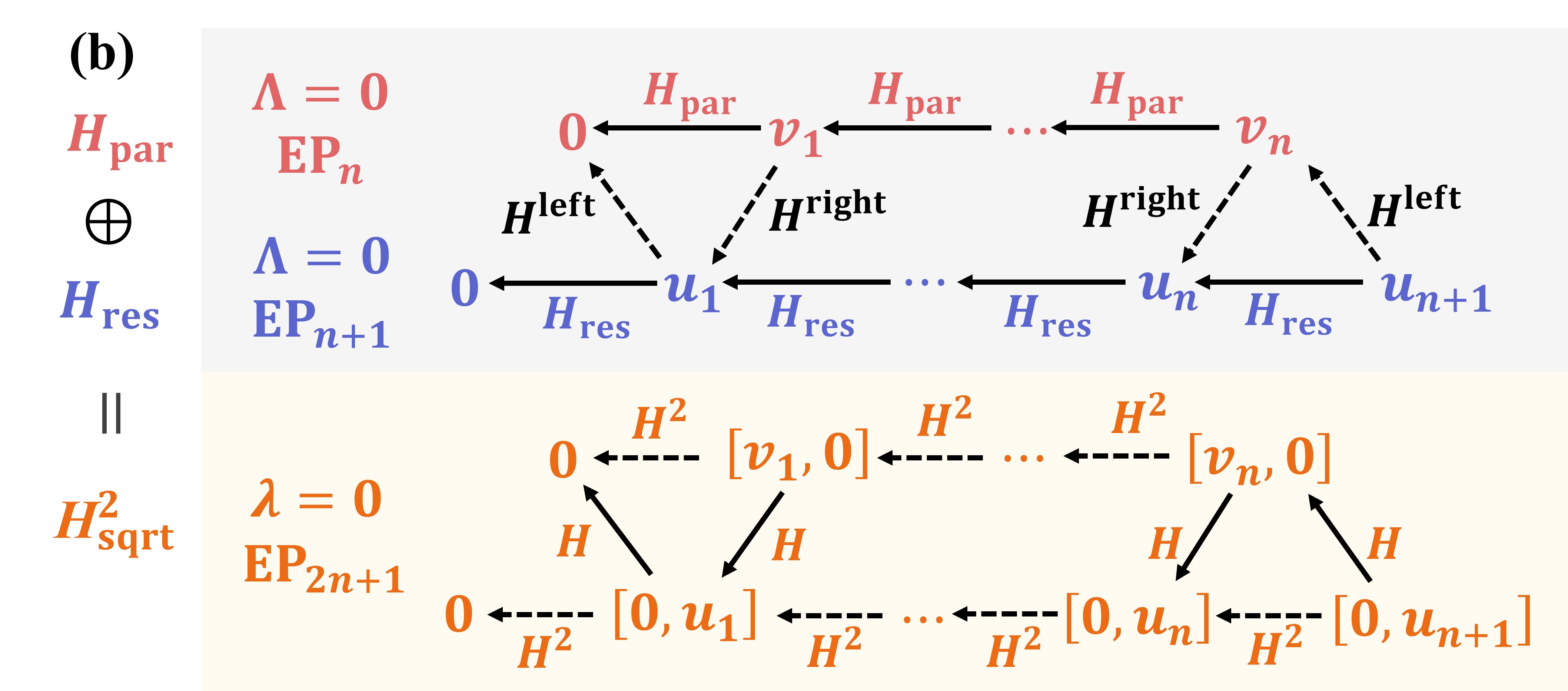}}
	\caption{Jordan chain structure of the parent, residual, and square-root systems  with parent EP$_n$ located at (a) $\Lambda\neq0$ and (b) $\Lambda=0$.}
	\label{fig:2}
\end{figure*}

By exploiting the connection between the square-root  and the parent Hamiltonian, the eigenvalues and eigenstates of the square-root system can be obtained.
According to Eq.~\eqref{eq:2}, the square-root system hosts two eigenvalues $\lambda_\pm=\pm\sqrt{\Lambda}$ with two eigenstates  $[\pm\sqrt{\Lambda} \bm{v}, H_{\rm right}\bm{v}]^T$, where  $\bm{v}$ is the eigenvector of the parent system with eigenvalue $\Lambda$.
Based on this, we further analyze the emergence of higher order of EPs.
Mathematically, the structure of the coalescence eigenvectors at an EP system is described by the Jordan chain, the length of which determines the order of EP \cite{1995springerPerturbation}.
When the parent EP$_n$ occurs at a nonzero eigenvalue $\Lambda$, the square-root system will host two sets of EP$_n$ with eigenvalues $\pm\sqrt{\Lambda}$.
The Jordan chain  $\{\bm{v}_j\}_{j=1}^{n}$ of parent system satisfies the chain relation $(H_{\mathrm{par}}-\Lambda I)\bm{v}_j = v_{j-1}$ with $(H_{\mathrm{par}}-\Lambda I)\bm{v}_1 = 0$. 
Then,   the eigenvectors of the square-root system can be constructed as  $\bm{s}_i = (H_{\rm sqrt}+\sqrt{\Lambda}I)^i [\bm{v}_i,\,0]^{\mathrm{T}}$ and $\bm{h}_i = (H_{\rm sqrt}-\sqrt{\Lambda}I)^i [\bm{v}_i,\,0]^{\mathrm{T}}$. They satisfy the Jordan-chain relations of the square-root system, corresponding to two sets of EP$_n$ located at $\lambda_{\pm} = \pm\sqrt{\Lambda}$ [Fig.~\ref{fig:2}(a)]. Also, the residual system has a Jordan chain $\{\bm{u}_j\}_{j=1}^{n}$ with the same EP order as that of parent system  \cite{SM}. 
Thus, starting from a generic parent system with EP$_n$, its square-root system can host two EP$_n$.

Then, we demonstrate that when the parent EP$_n$ occurs at zero energy, the two EP$_n$ in the square-root system coalesce into a single EP$_{2n+1}$. The Jordan chain of the square-root system is rearranged as an alternating sequence  of Jordan-chain vectors of parent and residual systems  [Fig.~\ref{fig:2}(b)]. The square-root Jordan chain is given by
\begin{equation}
	\begin{aligned}
		r_{2j-1}&=
		\begin{pmatrix}
			\textbf{0}\\ \bm{u}_j
		\end{pmatrix},
		\quad j=1,\ldots,n+1,
		\\
		r_{2j}&=
		\begin{pmatrix}
			\bm{v}_j\\ \textbf{0}
		\end{pmatrix},
		\quad j=1,\ldots,n,
	\end{aligned}
	\label{eq:3}
\end{equation}
where $\{\bm{v}_i\}_{i=1}^n$ and $\{\bm{u}_i\}_{i=1}^{n+1}$ denote the Jordan-chain vectors of the parent and residual systems, respectively. Eq.~\eqref{eq:3} satisfies that $H_{\mathrm{sqrt}}\,\bm{r}_k = \bm{r}_{k-1}$ for $k = 2, \ldots, 2n+1$ and $H_{\mathrm{sqrt}}\,\bm{r}_1 = 0$.
As a result, its chain length equals to the sum of the parent and residual chain lengths, namely, the chain length $2n+1$ is the order of square-root EP.
In particular, when the parent EP$_n$ is at zero energy, to support the formation of an EP$_{2n+1}$ in the square-root system,  the residual system hosts an EP$_{n+1}$ \cite{SM}. 
Note that the square-root system still has an EP$_{2n+1}$ even if the parent EP$_n$ is not located in zero energy, because the energy reference can be selected in our procedure. 


\begin{table*}               
	\caption{Symmetry properties and response orders for different perturbation classes.}
	\renewcommand{\arraystretch}{1.8}  
	\label{tab:1}        
	\begin{ruledtabular}       
		\begin{tabular}{l l l l l l} 
			Perturb. term & Perturb. location & Perturb. Symm. & Spectral Symm. & Max index separation & Response order \\\hline
			$V^{\rm par}$  & Parent sublattice  & $\Gamma V\Gamma=V$    & Broken  & $(v_1,0)\leftrightarrow(v_n,0)$   & $2n-1$  \\\hline
			$V^{\rm left}$ and $V^{\rm right}$  & \makecell[l]{Parent-residual\\coupling}  & $\Gamma V\Gamma=-V$    & Preserved  & \makecell[l]{$(0,u_1)\leftrightarrow(v_n,0)$\\$(v_1,0)\leftrightarrow(0,u_{n+1})$}   & $2n$\rule[-11pt]{0pt}{30pt}  \\\hline
			$V^{\rm res}$  & Residual sublattice  & $\Gamma V\Gamma=V$    & Broken  & $(0,u_1)\leftrightarrow(0,u_{n+1})$   & $2n+1$
		\end{tabular}
	\end{ruledtabular}
\end{table*}

\prlhead{Symmetry-induced multiple spectral responses.} 
The response order of HOEP systems is often associated with their intrinsic symmetry. 
While in previously studies on HOEP systems \cite{2014prlEnhancing,2017natureEnhanced}, owing to a lack of symmetry-guided design, they did not readily exhibit multiple response orders.
Here,  chiral symmetry is introduced by the square-root procedure, so the proposed HOEP system possesses a Jordan chain that alternates between two symmetry classes, i.e.,  between the parent and residual sites.
Mathematically, the response order is determined by the maximum separation of the Jordan-chain index that the perturbation bridges.
As a result, by applying perturbations at different lattice sites or at different couplings between sites, one can access different Jordan-chain lengths, thereby producing different response orders in HOEP systems.

In order to study the chiral symmetry of the square-root Hamiltonian, we define the chiral operator
\begin{equation}
\Gamma =
\begin{pmatrix}
I_{\mathrm{par}} & 0\\
0 & -I_{\mathrm{res}}
\end{pmatrix},
\end{equation}
which preserves the wave functions at the parent sites and inverts those at residual sites, corresponding to two eigenspaces with eigenvalues \(+1\) and \(-1\), respectively.
Since  \(H_{\mathrm{sqrt}}\) anticommutes with the chiral operator, i.e., \(\{H_{\mathrm{sqrt}}, \Gamma\} = 0\), it possesses chiral symmetry, which enforces spectral symmetry about zero energy [Fig.~\ref{fig:1}(b)].
Moreover, chiral symmetry imposes a distinctive structure on the generalized eigenspace of the square-root system, that is, the Jordan-chain vectors alternate between the parent and residual sites [Fig.~\ref{fig:2}(b)].

With above chiral-symmetric structure, we can analyze the response order of  this square-root HOEP system on the small perturbation.
Mathematically, the spectral response order is determined by the maximum Jordan-chain index separation bridged by the perturbation \cite{SM}.
A perturbation of strength $\epsilon$ that couples the Jordan-chain states $\bm{r}_i$ and $\bm{r}_j$ in Eq.~\eqref{eq:3} induces an energy splitting $\Delta E \propto \epsilon^{1/\nu},$ where $\nu=|i-j|+1$  corresponds to the response order of this perturbation.
Here the alternating Jordan-chain structure induced by chiral symmetry makes the bridged index interval location-dependent, so different perturbation locations yield different response orders.

Within the framework established above, we analyze the response properties of a specific perturbation $V$, given by
\begin{equation}
V=\left[\begin{array}{cc}{V}^{\rm par}& {V}^{\rm left}\\ {V}^{\rm right}& {V}^{\rm res}\end{array}\right].
\end{equation}
where ${V}^{\rm par}$ and $V^{\rm res}$ are perturbations within parent and residual sites, respectively, while ${V}^{\rm left}$ and ${V}^{\rm right}$ are perturbations between them.
Different perturbation locations correspond to different symmetry properties and response orders, as summarized in Table~\ref{tab:1}, where  response orders of  three classes are analyzed.

Class 1: A perturbation is applied only within parent sites, denoted as $V^{\rm par}$. It satisfies $\Gamma V\Gamma=V$, thereby breaking chiral symmetry and associated spectral symmetry.
It only couples even-indexed Jordan-chain states located in parent sites, so 
the largest index difference is \(2n-2\), attained by
$r_2\leftrightarrow r_{2n}$, with the maximal response order \(2n-1\).

Class 2: A perturbation is applied to the coupling between the parent and residual sites, denoted as $V^{\rm left}$ and $V^{\rm right}$. This class of perturbations satisfies $\Gamma V \Gamma = -V$,  thereby preserving chiral symmetry and associated spectral symmetry.
It only couples odd- and even-indexed Jordan chain states, so
the largest index difference is \(2n-1\), attained by
$r_1\leftrightarrow r_{2n}$ and $r_2\leftrightarrow r_{2n+1}$, with the maximal response
order \(2n\).

Class 3: A perturbation is applied only within residual sites, denoted as $V^{\rm res}$. it satisfies $\Gamma V\Gamma=V$ and thereby breaking chiral symmetry and associated spectral symmetry.
It only couples odd-indexed Jordan chain states located in residual sites, so the largest index difference is \(2n\), attained by
$r_1\leftrightarrow r_{2n+1}$, with the maximal response order \(2n+1\).

Therefore, based on above symmetric anylasis, we can explain why perturbations at different locations produce spectral responses with distinct orders. 
Notably, when the square-root operation is applied iteratively, the square-root system further acquires composite symmetries and exhibits multiple responses of even higher orders \cite{SM}. 
This mechanism enhances the richness of EP responses and thus holds promise for applications in signal amplification, sensing, and non-Hermitian control.

\begin{figure}[b]
	\includegraphics[width=\figsize\textwidth]{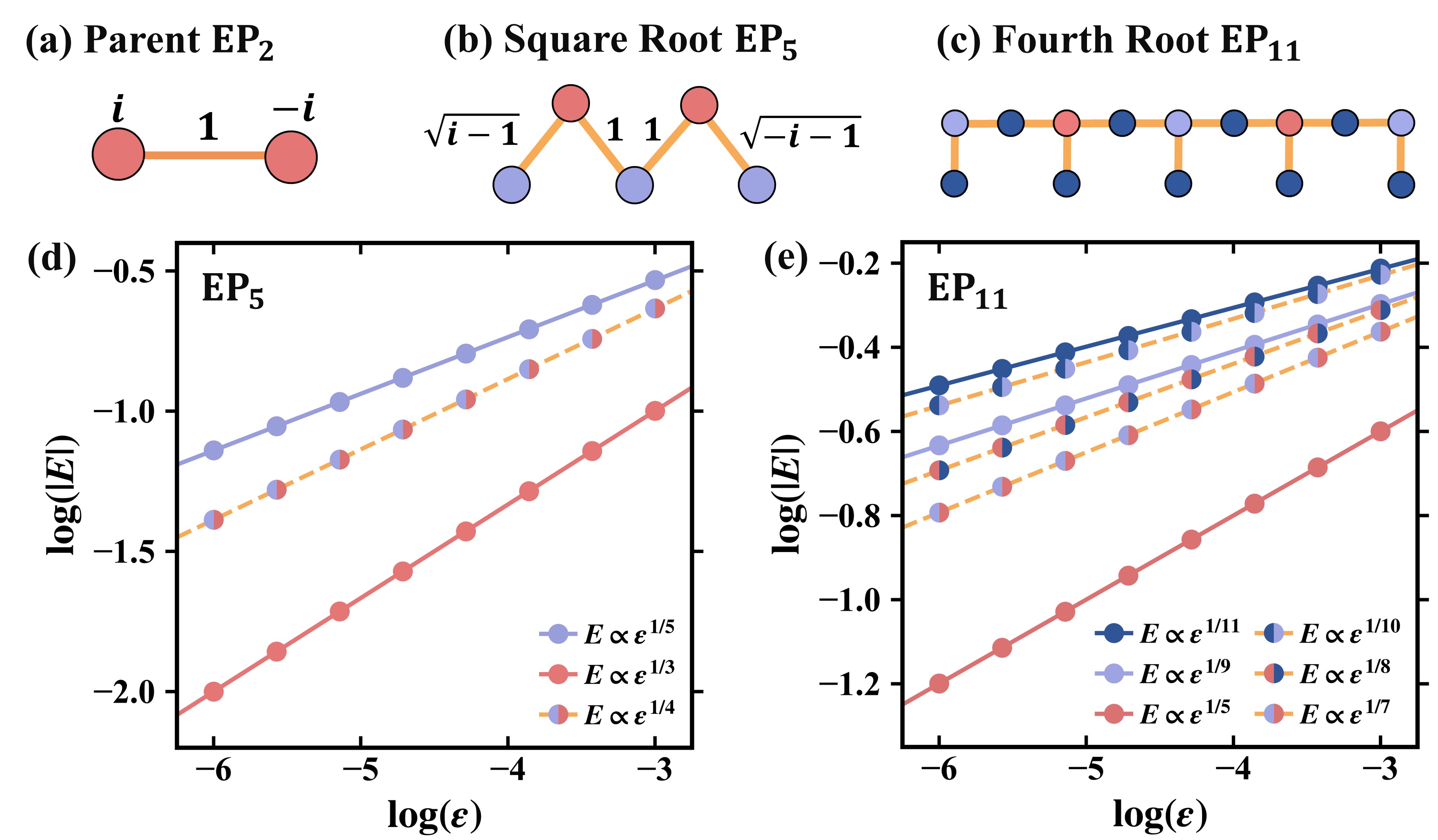}
	\caption{\label{fig:3}(a) PT-symmetric two-site EP$_{2}$ parent system. (b) EP$_{5}$ square-root system from square-rooting (a). (c) EP$_{11}$ square-root system from square-rooting (b). Eigenvalue splittings of (d) EP$_{5}$ and (e) EP$_{11}$ under perturbation $\epsilon$ imposed on parent sites, residual sites, and couplings.} 
\end{figure}

\prlhead{Iterative generation of HOEPs.}
We illustrate the iterative construction of HOEPs starting from an EP$_{2}$ parent system. 
The parent is a two-site tight-binding dimer with balanced gain/loss $\pm i\gamma$ and real hopping $\kappa$. At $\kappa=\gamma$, the parent Hamiltonian reads
\begin{equation}
	H_{\mathrm{par}}=\kappa \left(\begin{array}{cc}i& 1\\ 1& -i\end{array}\right).
\end{equation}
Its eigenvalues and eigenstates then coalesce at zero energy, realizing an EP$_{2}$ \cite{2023JoPGain-gain,2026prajump}. For simplicity, we set $\kappa=\gamma=1$ in the following, as shown in [Fig.~\ref{fig:3}(a)].
The square-root construction of this EP$_2$ parent system introduces three residual sites and yields an EP$_5$ square-root system [Fig.~\ref{fig:3}(b)], whose Hamiltonian is
\begin{equation}
	{H}_{\rm sqrt}=\left(\begin{array}{ccccc}0& 0& \sqrt{i-1}& 1& 0\\ 0& 0& 0& 1& \sqrt{-i-1}\\ \sqrt{i-1}& 0& 0& 0& 0\\ 1& 1& 0& 0& 0\\ 0& \sqrt{-i-1}& 0& 0& 0\end{array}\right).
\end{equation}
For definiteness, we take $\sqrt{i-1}$ as $2^{1/4}e^{3\pi i/8}$, and $\sqrt{-i-1}$ as $2^{1/4}e^{5\pi i/8}$, without affecting the following analysis.
This square-root system has a unique eigenstate $(0,0,e^{5\pi i/8},2^{1/4},e^{3\pi i/8})$ at zero energy and therefore realizes an EP$_5$. For perturbations applied to parent and residual sites, the response orders are 3 and 5, respectively. While, for a perturbation that couples parent and residual sites, the response order is 4. The dependence of a perturbed eigenvalue magnitude $|E|$ on the perturbation strength $\epsilon$, applied at different locations, is shown in Fig.~\ref{fig:3}(d).

Applying the square-root operation once more to the resulting EP$_5$ system yields an EP$_{11}$ system [Fig.~\ref{fig:3}(c)].
This second-generation square-root system possesses multiple chiral symmetries, which force each Jordan-chain vector in its length-11 chain to be supported entirely on one of three site classes, namely, the original EP$_2$ parent sites (red), the first-generation residual sites (light-blue), and the second-generation residual sites (dark-blue) \cite{SM}.
Consequently, this EP$_{11}$ system exhibits additional response orders beyond the three classes identified in the previous section.
Specifically, perturbations applied to the original parent sites, the first-generation residual sites, and the second-generation residual sites produce 5$^{\rm th}$-, 9$^{\rm th}$-, and 11$^{\rm th}$-order responses, respectively.
Perturbations of the couplings between different site classes produce \(7^{\rm th}\)-, \(8^{\rm th}\)-, and \(10^{\rm th}\)-order responses, with the correspondence specified in Fig.~\ref{fig:3}(e).

This procedure can be continued iteratively, yielding EP$_{23}$, EP$_{47}$, and so forth.
This implies that the square-root method can be applied iteratively to construct EP systems of arbitrarily high order, demonstrating the scalability and broad applicability of the proposed construction.

\prlhead{Conclusions.}
We extend the square-root operation to finite-lattice non-Hermitian systems, providing a spectral-engineering framework applicable to arbitrary tight-binding models.
Starting from a parent Hamiltonian hosting an $\mathrm{EP}_n$, the square-root construction yields a chiral-symmetric system hosting an $\mathrm{EP}_{2n+1}$.
The proposed scheme is universal, which can be applicable to any existing tight-binding EP system and iterable to generate EPs of arbitrarily high order.
Note that even from a parent system that does not possess EP, we can also generate an HOEP square-root system, which further demonstrates the generality of our method.
Symmetry analysis shows that perturbations applied at different locations induce energy splittings of different orders, demonstrating that symmetry-based construction can increase the EP order and broaden the range of accessible response orders.
The resulting systems may enable sensitive measurements and controlled light--matter interactions, with potential applications in sensing and non-Hermitian control.

\begin{acknowledgments}
	This work is supported by the National Natural Science Foundation of China under Grant No. 12474370 and No. U25D9003 and the Quantum Science and Technology-National Science and Technology Major Project No. 2021ZD0301500.
\end{acknowledgments}


\end{document}